\documentclass[aps,twocolumn,nofootinbib]{revtex4}
\usepackage[latin1]{inputenc}
\usepackage{epsfig}
\usepackage{graphicx}
\usepackage{amsfonts}
\usepackage{amssymb}
\usepackage{calrsfs}
\usepackage{frcursive}
\newcommand{\beq}{\begin{equation}}
\newcommand{\eeq}{\end{equation}}
\newcommand{\la}{\langle}
\newcommand{\ra}{\rangle}
\newcommand{\bxi}{\mbox{\boldmath $\xi$}}
\newcommand{\bzeta}{\mbox{\boldmath $\zeta$}}
\newcommand{\bomega}{\mbox{\boldmath $\omega$}}
\newcommand{\bx}{\bf x}
\newcommand{\bp}{\bf p}
\newcommand{\bv}{\bf v}
\newcommand{\bk}{\bf k}
\newcommand{\bq}{\bf q}
\newcommand{\ba}{\bf a}
\newcommand{\bb}{\bf b}

\begin{document}

\title{Stochastic motion in phase space
on a surface of constant energy}

\author{Tânia Tomé and Mário J. de Oliveira}
\affiliation{Universidade de São Paulo, 
Instituto de Física, Rua do Matão, 1371,
05508-090 São Paulo, SP, Brasil.}

\begin{abstract}

We study closed systems of particles that are subject to
stochastic forces in addition to the conservative forces.
The stochastic equations of motion are set up in such a
way that the energy is strictly conserved at all times.
To ensure this conservation law, the evolution equation
for the probability density is derived using an appropriate
interpretation of the stochastic equation of motion that
is not the Itô nor the Stratonovic interpretation. 
The trajectories in phase space are restricted
to the surface of constant energy. Despite this
restriction, the entropy is shown to increase with
time, expressing irreversible behavior and
relaxation to equilibrium. This main result of the
present approach contrasts with that given by the
Liouville equation, which also describes closed
systems, but does not show irreversibility.

\end{abstract}

\maketitle

\section{Introduction}

One of the fundamental problems of non-equilibrium
statistical mechanics is the explanation of the
irreversible decay to equilibrium of closed systems,
which is expressed by the increase of entropy. If we
consider a closed system of particles interacting
through conservative forces, then we are naturally
led to a description of the system through the
Liouville equation
\cite{gibbs1902,borel1925,tolman1938,khinchin1949,
landau1958,prigogine1962,hobson1971,zubarev1974,
arnold1978,reichl1980,salinas2001}.
However, a direct result of this equation
is the invariance of entropy with time and as
a consequence the equation is unable to describe
the desired irreversible decay to equilibrium.

The Liouville equation describes appropriately
systems in equilibrium. Any probability density
$\rho$ which depends on the dynamic variables
through the Hamiltonian ${\cal H}$ is a stationary
solution of this equation. This is the case of the
microcanonical Gibbs distribution
\beq
\rho_e = \frac1\Omega \delta(E-{\cal H}),
\label{9}
\eeq
which describes a system in thermodynamic equilibrium 
with a fixed energy $E$ \cite{gibbs1902}. A relevant
property of the Liouville equation is the conservation
of the energy. If the initial probability density is
defined on a surface of a constant given energy,
then the density will remain forever defined
on this surface. 

The Liouville equation predicts 
that the entropy remains invariant in time, a result which
is a direct consequence of the constance of the probability
density along a trajectory in phase space. If we start with
a probability density with an entropy distinct from that
of the Gibbs equilibrium distribution, we conclude
that this distribution will never be reached and that the
Liouville equation does not account for the irreversibility.
This inconvenience may be circumvent by introducing random
forces that changes the Hamiltonian trajectory while keeping
the conservation of energy and momentum.

The Boltzmann kinetc equation
\cite{huang1963,rumer1980,cercignani1988,
kremer2010,oliveira2019} 
is in fact a realization of this
proposal. It incorporates random forces and as a consequence
the entropy increases while the energy remains strictly
constant. Here we derive a stochastic equation that is
similar to Boltzmann equation in the sense that the entropy
also increases at strictly constant energy. 
To this end we set up a stochastic equation of motion,
or Langevin equations, that strictly conserves the energy,
that is the trajectories in phase space lie
on the surface of constant energy. 
From the stochastic equation of motion one derives
the associated evolution equation for the probability
density, which turns up to be an equation of the
Fokker-Planck type
\cite{kampen1981,risken1989,gardiner2009,tome2015},
and it is not an integro-differential equation as is the
case of the Boltzmann equation.

The time evolution equation for the probability density,
which incorporates stochastic forces, is a type of
equations that is used in approaches to stochastic thermodynamics 
\cite{tome2010,vandebroeck2010,spinney2012,seifert2012,
santillan2013,luposchainsky2013,tome2015a,oliveira2020,oliveira2020a},
which describe systems in contact with heat reservoirs,
including the exchange of heat and entropy as well as the
production of entropy.
The main difference of these approaches from ours is
the strictly conservation of energy observed in the
present approach. In this sense
we may say that the present approach provides 
a stochastic thermodynamics for closed systems.

To derive the evolution equation, it is necessary to
give an interpretation to the stochastic equation of motion because
the noise is multiplicative \cite{kampen1981}. Usually,
the interpretation is that given by Stratonovich
and not that given by Itô. However, we find that
the proper interpretation necessary to preserve
energy at all times is none of them. 

The origin of stochastic trajectories may be attributed
to forces that are of the hyperbolic type such as those
occurring when two hard spheres collide with each other.
Given the initial positions and velocities of two hard
spheres, then the positions and velocities after the
collision will be uniquely determined by the Newton
equations of motion. However, if we consider two very
similar initial conditions, that differ only slightly
by the positions, the velocities being the same, then 
the velocities after the collision will not differ
slightly but can be very different. Since the initial
condition is almost the same, this result looks as if
more than one trajectory is possible from a single
initial condition.

We will not pursue in more detail the problem of how 
stochastic trajectories emerge from pure mechanics,
that is, from the Newton equations of motion applied
to a system acted by internal conservative forces.
Here we adopt the point of view that the stochastic
trajectories or rather the stochastic forces are an
assumption of the present approach.

The stochastic equation of motion that we have introduced
can be used as a numerical method to simulate a system of
interacting particles at constant energy, as is the case
of the method of molecular dynamics
\cite{allen1987,tuckerman2010}. 
In equilibrium, the static properties will be the same 
as those obtained from the molecular dynamics because
the stochastic equations of motion lead to the Gibbs
distribution. However, the two-time correlations
will be distinct due to presence of the stochastic forces.

\section{Impulsive stochastic force}

\subsection{Stochastic equation of motion}

We consider two particles interacting through random
forces that act on a short period of time. 
We denote by ${\bp}_1=m_1 {\bv}_1$ and ${\bp}_2=m_2{\bv}_2$
the momenta of the particles, where $m_1$ and $m_2$
are their masses and $\bv_1$ and $\bv_2$
their velocities. The random force acting on particle
1 due to the particle 2 is denoted by $\bxi$.
The reaction on particle 2 is $-\bxi$ and the
equations of motion are
\beq
\frac{d{\bp}_1}{dt} = {\bxi},
\qquad
\frac{d{\bp}_2}{dt} = - {\bxi}.
\label{69}
\eeq 
The total momentum ${\bf P}=\bp_1+\bp_2$ is conserved
because from the stochastic equation of motion it follows
that $d{\bf P}/dt=0$. 

We wish to determine the properties of the stochastic
vector variable $\bxi$ that makes the energy of
the two particles constant. As the forces are suppose
to be impulsive, which means that they act during
a short period of time, the conservation of energy
means the conservation of the kinetic energy
\beq
{\cal K} = \frac{p_1^2}{2m_1} + \frac{p_2^2}{2m_2}.
\eeq 
The derivative of the kinetic energy is
\beq
\frac{d{\cal K}}{dt} = ({\bv}_1 - {\bv}_2) \cdot {\bxi},
\eeq
Defining the relative velocity ${\bv}={\bv}_1-{\bv}_2$,
we see that the condition 
\beq
{\bv} \cdot {\bxi} = 0
\label{15}
\eeq
makes the derivative of the kinetic energy to vanish.
This condition of orthogonality implies that
$\bxi$ depend on the relative velocity $\bv$.

To proceed further we use a more convenient
procedure. As $\bp_1$ and $\bp_2$ are connected
by the conservation of the total momentum,
if suffices to use just one independent variable
which we choose to be
${\bp}=m{\bv}$ where $m$ is the reduced mass.
The two equations of motion are then reduced to one,
\beq
\frac{d{\bp}}{dt} = {\bxi}.
\qquad
\label{70}
\eeq 
In terms of the new variable ${\bp}$, the kinetic energy,
apart from a constant, is given by
\beq
{\cal K} = \frac{p^2}{2m}.
\eeq

The stochastic variable $\bxi$ is understood
as follows. If we discretize
the time in intervals equal to $\tau$ the stochastic
variable is proportional to $1/\sqrt{\tau}$. Thus
in a discretize formulation of the equation of
motion (\ref{70}), we replace $\bxi$ by
${\bxi}/\sqrt{\tau}$ with the understanding
that $\bxi$ has a finite variance. 
The equation (\ref{70}) becomes
\beq
\Delta{\bp} = \sqrt{\tau} \bxi.
\label{71}
\eeq
A finite increment of the kinetic energy is
\beq
\Delta{\cal K} = {\bv} \cdot \Delta{\bp}
+ \frac1{2m} (\Delta{\bp})^2.
\label{72}
\eeq
Replacing (\ref{71}) in this equation, we find
\beq
\Delta{\cal K} = \sqrt{\tau} {\bv\cdot\bxi}
+ \frac{\tau}{2m} {\bxi}\cdot{\bxi}, 
\eeq
Using the condition (\ref{15}) on $\bxi$  
established above, ${\bv} \cdot {\bxi} = 0$,
we see that $\Delta{\cal K}$ vanishes up to order
$\sqrt{\tau}$, but not up to order $\tau$.
To overcome this inconvenience, we observe that
(\ref{71}) is not the only possible interpretation
of the stochastic equation of motion (\ref{70}).
We may add any random variable proportional to
$1/\tau$ which means to interpret the stochastic 
equation of motion as
\beq
\Delta{\bp} = \sqrt{\tau} \bxi + \tau {\bomega}.
\label{71a}
\eeq
where ${\bomega}$ is a random variable with a
finite variance.
Replacing this expression in (\ref{72}), then
up to terms of order $\tau$ we find 
\beq
\Delta{\cal K} =  \sqrt{\tau} {\bv}\cdot \bxi
+ \tau ({\bv}\cdot {\bomega}
+ \frac{1}{2m} {\bxi}\cdot{\bxi}). 
\label{73}
\eeq

The first term vanishes due to the condition 
(\ref{15}). If we wish $\Delta{\cal K}$ to 
vanish to  order $\tau$, the last term must vanish,
which gives the following relation  
\beq
 {\bomega}\cdot {\bp}
+ \frac{1}{2} {\bxi}\cdot{\bxi} = 0,
\eeq
between $\bomega$ and $\bxi$. 
To find $\bomega$ in terms of $\bxi$, we differentiate the
condition (\ref{15}) with respect to ${\bp}$, and 
perform a dot product with $\bxi$ to obtain
\beq
{\bxi}\cdot(\frac{\partial}{\partial{\bp}} {\bxi})
\cdot {\bp} + {\bxi}\cdot{\bxi}= 0.
\eeq
Comparing the last two equation, we see that
the expression
\beq
{\bomega}'=\frac12{\bxi}\cdot
\frac{\partial}{\partial{\bp}} {\bxi},
\label{23}
\eeq
gives the desired solution. However, this is not
the unique solution because we may add to
this expression any term orthogonal to ${\bp}$
such as $\bxi$ itself. At this point 
we postpone the problem of determining
$\bomega$ in terms of $\bxi$.

\subsection{Evolution equation}

To derive the Fokker-Planck equation associated
to the Langevin equation (\ref{71a}), which is the
evolution equation for the probability density $\rho({\bp})$,
we use a method that consists in determining the evolution
of the characteristic function $\Phi$, which is
the Fourier transform of $\rho$,
\beq
\Phi = \int e^{i\Omega} \rho\, d{\bp},
\eeq
where $\Omega=i{\bk}\cdot{\bp}$ and ${\bk}$ is a vector
in the Fourier space.
The characteristic function is also the average 
\beq
\Phi = \la e^{i\Omega}\ra.
\eeq

Denoting by $\Delta \Phi$ the difference of the characteristic
functions between times $t+\tau$ and $t$, then
\beq
\Delta\Phi = \la e^{i\Omega}(e^{i{\bk}\cdot\Delta{\bp}}-1)\ra,
\eeq
where the average is to be calculated using the probability
of the variables $\bp$ and $\bxi$ considered to be independent
of each other.
Next we replace $\Delta{\bp}$, given by (\ref{71a}) in this
expression and expand $\Delta\Phi$ up to terms of order $\tau$.
Dividing the result by $\tau$, we find the time
derivative of the characteristic function in the form
\beq
\frac{\partial\Phi}{\partial t}
= \la e^{i\Omega} i{\bk}\cdot {\bomega} \ra
+ \frac12 \la e^{i\Omega} i{\bk}\cdot {\bxi}{\bxi}
\cdot i {\bk} \ra.
\label{80}
\eeq
Taking into account that 
$i{\bk}\cdot e^{i\Omega}=\partial e^{i\Omega}/\partial{\bp}$,
the first term on the right-hand side of (\ref{80})
is written as
\beq
\int (\frac{\partial e^{i\Omega}}{\partial{\bp}})
\cdot \la{\bomega}\ra\, \rho\, d{\bp}
= - \int e^{i\Omega} \frac{\partial }{\partial{\bp}}
\cdot \la{\bomega}\ra\, \rho\, d{\bp},
\eeq
where we have performed an integration by parts,
and now the average is on $\bxi$ only.
This expression is the Fourier transform of
\beq
- \frac{\partial }{\partial{\bp}}
\cdot \la{\bomega}\ra \rho.
\label{74b}
\eeq
In this and in other integration by parts, we are
considering that the integrated part disappears
by the assumption of a rapid vanishing of $\rho$
at the boundary of integration.

Using the same procedure, the second term on the
right-hand side of (\ref{80}) is written as
\beq
\frac12 \int e^{i\Omega} \frac{\partial }{\partial{\bp}}
\cdot \frac{\partial }{\partial{\bp}} \cdot  
\la{\bxi \bxi}\ra \rho\, d{\bp},
\eeq
where we have performed two successive integrations by
parts. This expression is the Fourier transform of
\beq
\frac12\frac{\partial }{\partial{\bp}} \cdot
\frac{\partial }{\partial{\bp}} \cdot  
\la{\bxi \bxi}\ra \rho.
\label{74c}
\eeq

To determine $\partial\rho/\partial t$, we bear in
mind that $\partial\Phi/\partial t$, given by
(\ref{80}), is its Fourier transform. Therefore,
the time derivative of $\rho$ is obtained by
adding up the expressions (\ref{74b}) and
(\ref{74c}). The result is
\beq
\frac{\partial\rho}{\partial t} = 
\frac{1}2 \frac{\partial }{\partial{\bp}} \cdot\left(
\frac{\partial }{\partial{\bp}} \cdot  
\la{\bxi \bxi}\ra \rho - 2\la{\bomega}\ra \rho \right).
\label{81}
\eeq
It remains yet to determine $\bomega$ in terms
of $\bxi$.

\subsection{Lagrange multiplier}

We have seen that the discretized equation of motion
(\ref{71a}) preserves the kinetic energy up to time
$\tau$. It is not guaranteed that the kinetic energy
${\cal K}$ will be preserved forever. To ensure that this will
happen for all times, we impose the vanishing of the
probability of a trajectory that do not lie on a
surface of constant kinetic energy. This amounts to
say that if initially $\rho$ is nonzero only on the
surface of constant ${\cal K}$, then this property should
be preserved as $\rho$ evolves in time. Let us denote
by $\rho^*$ a probability density that is zero outside
this surface. Then the variation of $\rho^*$ to a point
outside the surface vanishes, that is $d\rho^*=0$.
If $\rho$ is a generic solution of (\ref{81}), then
$\rho^*$ can be found from $\rho$ by using the method
of the Lagrange multipliers, which means
\beq
d\rho^* = d\rho + \lambda d{\cal K},
\eeq
where $\lambda$ is a Lagrange multiplier.
This equation is equivalent to
\beq
\rho^* = \rho + \lambda{\cal K}.
\eeq
Replacing this expression in the evolution equation
(\ref{81}), written in the simplified form
\beq
\frac{\partial\rho}{\partial t} = {\mathfrak F}\rho,
\label{87}
\eeq
where ${\mathfrak F}$ is a linear differential operator,
we find that $\rho^*$ is a solution of the evolution
equation as long as
\beq 
{\mathfrak F}{\cal K} = 0.
\label{85}
\eeq

The condition (\ref{85}) becomes fulfilled if
the expression between parentheses in 
(\ref{81}) vanishes when $\rho$ is replaced by
$K=p^2/2m$, that is,
\beq
\frac{\partial }{\partial{\bp}} \cdot  
{\bxi \bxi} p^2
- 2{\bomega} p^2 = 0.
\eeq
Solving for $\bomega$, we find
\beq
{\bomega} =\frac1{2p^2}\frac{\partial }{\partial{\bp}}
\cdot {\bxi \bxi} p^2,
\eeq
which, by the use of the relation
${\bp}\cdot{\bxi}=0$, can be written in the more
simplified form
\beq
{\bomega} =\frac12\frac{\partial }{\partial{\bp}}
\cdot {\bxi \bxi}.
\label{83}
\eeq
which is the sought relation between $\bomega$ and
$\bxi$.

Replacing (\ref{83}) in (\ref{81}) we obtain the 
evolution equation in the simple form
\beq
\frac{\partial\rho}{\partial t} = 
\frac{1}2 \frac{\partial }{\partial{\bp}} \cdot
\la{\bxi \bxi}\ra\cdot
\frac{\partial }{\partial{\bp}}\rho.
\label{81a}
\eeq
Using (\ref{83}), the discretized stochastic equation
of motion (\ref{71a}) becomes
\beq
\Delta{\bp} = \sqrt{\tau} \bxi + 
\frac\tau2\frac{\partial }{\partial{\bp}}
\cdot {\bxi \bxi},
\label{71b}
\eeq
which is the desired interpretation of the stochastic
equation of motion leading to the conservation of
kinetic energy for all times.
It is worth mentioning that this equation
{\it does not} correspond to the Stratonovich 
interpretation of (\ref{70}), which is
\beq
\Delta{\bp} = \sqrt{\tau} \bxi + 
\frac\tau2 {\bxi}\cdot\frac{\partial}{\partial{\bp}} {\bxi},
\label{71c}
\eeq
and corresponds to use ${\bomega}'$, given by (\ref{23}).
Of course, it does not correspond either to the Itô 
interpretation which is simply that given by
(\ref{71}).

\subsection{Expression of $\bxi$}

It remains now to determine the explicit expression
for the stochastic force $\bxi$. The only restriction
that has to be fulfilled is the orthogonal
condition (\ref{15}), that is,
\beq
{\bv} \cdot {\bxi} = 0.
\label{15d}
\eeq
To meet this condition, we choose two
vectors ${\ba}$ and $\bb$ that are orthogonal 
to $\bv$, and to each other, 
\beq
{\bv} \cdot {\ba} = 0, \qquad {\bv} \cdot {\bb} = 0,
\qquad {\ba} \cdot {\bb} = 0,
\label{15e}
\eeq
and write
\beq
{\bxi} = {\ba}\sigma + {\bb} \eta,
\eeq
where $\sigma$ and $\eta$ are independent 
stochastic variable with zero mean and variance
equal to $2\gamma$.

From these definitions we determine the
average $\la{\bxi\bxi}\ra$ appearing in the
evolution equation (\ref{81a})
\beq
\la{\bxi\bxi}\ra = 2\gamma({\ba} {\ba} + {\bb}{\bb}),
\eeq
which replaced in (\ref{81}), gives
\beq
\frac{\partial\rho}{\partial t} = \gamma
\frac{\partial }{\partial{\bp}} \cdot
({\ba} {\ba} + {\bb}{\bb}) \cdot
\frac{\partial }{\partial{\bp}}\rho.
\label{81b}
\eeq

We write the Cartesian coordinates of the relative
velocity $\bv$ using spherical coordinates as
\beq
{\bv} = (v\sin\theta \cos\varphi,\,\,
v\sin\theta \sin\varphi,\,\, v\cos\theta),
\eeq
where $v$ is the absolute value of the velocity,
$\theta$ is the polar angle and $\varphi$ is the
azimuthal angle. The vectors $\ba$ and $\bb$ are
chosen to be unit vector with Cartesian coordinates
given by
\beq
{\ba} = (\cos\theta\cos\varphi,\,\,
\cos\theta\sin\varphi,\,\, -\sin\theta),
\label{25a}
\eeq
\beq
{\bb} = (-\sin\theta\sin\varphi, \,\,
\sin\theta\cos\varphi,\,\, 0),
\label{25b}
\eeq
It is easily seen that the conditions (\ref{15e})
are satisfied.

Up to now we have treated a system in three dimensions. 
To treat a two-dimensional system, we have to consider
that the vector quantities have two Cartesian
components. In this case, instead of two vectors
orthogonal to ${\bf v}$, just one is possible.
Writing 
\beq
{\bv} = (v\cos\phi,\,\, v\sin\phi).
\eeq
this vector is
\beq
{\bb} = (-\sin\phi,\,\, \cos\phi),
\eeq
In this case there is just one random variable
$\eta$ and
\beq
{\bxi} = {\bb} \eta.
\eeq

If we wish to treat a system in one dimension,
we see that this is unattainable because it is
not possible to meet the condition of
orthogonality (\ref{15e}). This reflects the
following result concerning the collision of
two particles in one dimension. Given the
velocities of the particles before the collision,  
they are uniquely determined after the collision,
if the energy and momentum are conserved. Thus in
one dimension there is no room for a stochastic
motion that conserves both the energy and momentum.

\subsection{Entropy production}

Although the energy is strictly conserved, thist is
not the case of the entropy. The entropy is defined
by 
\beq
S = - k \int \rho\ln\rho\, d{\bp},
\eeq
and its time variation can be obtained from the
the evolution equation (\ref{81b}). Deriving this
expression with respect to time, we obtain
\beq
\frac{dS}{dt} = - k \int \frac{\partial\rho}{\partial t}
\ln\rho\, d{\bp}.
\eeq
Replacing the derivative of $\rho$, given by the
evolution equation (\ref{81b}), and after an
integration by parts, we reach the result
\beq
\frac{dS}{dt} = k \gamma \int \frac1\rho 
\frac{\partial\rho}{\partial{\bp}} 
\cdot
({\ba} {\ba} + {\bb}{\bb}) \cdot
\frac{\partial\rho}{\partial{\bp}}d{\bp},
\eeq
which can be written in the form
\beq
\frac{dS}{dt} = k \gamma \int \frac1\rho 
(A^2 + B^2) d{\bp},
\label{76}
\eeq
where
\beq
A = {\ba}\cdot\frac{\partial\rho}{\partial{\bp}},
\qquad
B = {\bb}\cdot\frac{\partial\rho}{\partial{\bp}}.
\eeq
As the integral is positive definite, $dS\geq0$,
the entropy is a monotonic increasing function
of time. The right-hand side of (\ref{76}) is
understood as the rate of entropy production.
In the stationary state $\rho_e$ will be
a function of $K$ and $\partial\rho/\partial{\bp}$
will be proportional to ${\bp}$ which is orthogonal
to $\ba$ and $\bb$. Thus $A$ and $B$ vanish, the
entropy production vanishes and
the entropy reaches is maximum value.

\section{System of interacting particles}

\subsection{Stochastic equations of motion}

Our attention is now directed toward a system of
several interacting particles. The position and
the momentum of particle $i$ are denoted by
${\bf x}_i$ and by ${\bp}_i$, and the system
is described by the Hamiltonian function
\beq
{\cal H} = \sum_i \frac{p_i^2}{2m_i} + {\cal V},
\label{1}
\eeq
where $p_1^2=p_{xi}^2+p_{yi}^2+p_{zi}^2$,
and $m_i$ is the mass of the particle $i$,
and ${\cal V}$ is a function of the coordinates
and represent the potential energy of the particles.
The stochastic equations of motion, or the Langevin
equations, are
\beq
\frac{d{\bx}_i}{dt} = {\bv}_i,
\qquad\qquad
\frac{d{\bp}_i}{dt} = {\bf F}_i + {\bzeta}_i,
\label{5}
\eeq
where ${\bv}_i$ and ${\bf F}_i$ are the velocity
and the conservative force, respectively,
associated to the particle $i$, and are given by
\beq
{\bv}_i = \frac{\partial{\cal H}}{\partial{\bp}_i},
\qquad\qquad
{\bf F}_i = - \frac{\partial{\cal H}}{\partial{\bx}_i}.
\eeq

We wish to describe a system by forces that
strictly conserves the total momentum 
of the collection of particles, given by
\beq
{\bf P} = \sum_i {\bp}_i,
\eeq
and the total energy, given by (\ref{1}).
The particles are subject only to internal 
forces which means that for each force ${\bf F}_i$ 
there is a reaction force with the opposite sign so
that the sum of the conservative forces vanish,
\beq
\sum_i {\bf F}_i = 0.
\label{19}
\eeq
Deriving ${\bf P}$ with respect to time and using
(\ref{19}), we get
\beq
\frac{d{\bf P}}{dt} = \sum_i {\bzeta}_i.
\label{6a}
\eeq
Therefore the conservation of momentum requires
that the sum of the stochastic forces vanishes.

Deriving the function ${\cal H}$ with respect
to time, we find
\beq
\frac{d{\cal H}}{dt} = \sum_i {\bv}_i \cdot \bzeta_i.
\label{6b}
\eeq
There is no contribution coming from the
conservative forces. The right-hand side of this
equation is the total power of the stochastic force,
which should vanish.

The stochastic forces are chosen so that the right-hand
sides of the equations (\ref{6a}) and (\ref{6b})
vanish identically. To meet the first requirement,
we choose ${\bzeta}_i$ as a sum of independent
stochastic vector variables ${\bxi}_{ij}$,
\beq
{\bzeta}_i = \sum_j {\bxi}_{ij}.
\label{7}
\eeq
with the properties
\beq
{\bxi}_{ji} = -{\bxi}_{ij},
\label{13}
\eeq
and ${\bxi}_{ii}=0$. We see that the right-hand side of
(\ref{6a}) vanishes identically, and the total momentum
is strictly conserved. The vector ${\bxi}_{ij}$ is
understood as the random force acting on particle $i$ due to
particle $j$ and ${\bxi}_{ji}$ as the random force on
particle $j$ due particle $i$. Since they were chosen
to differ only by their signs, they are interpreted as
action and reaction, leading to the conservation of the
total momentum. We now replace (\ref{7}) in the
right-hand side of equation (\ref{6b}) to find
\beq
\frac{d{\cal H}}{dt} = \frac12\sum_{ij}
({\bv}_i - {\bv}_j) \cdot {\bxi}_{ij},
\label{6c}
\eeq
where we have used the property (\ref{13}).
Requiring that 
\beq
{\bv}_{ij} \cdot {\bxi}_{ij} = 0,
\label{8}
\eeq
for each pair $ij$, 
where ${\bv}_{ij}={\bv}_i - {\bv}_j$
is the relative velocity of particles $i$ and $j$,
we see that the right-hand side of (\ref{6c})
vanishes identically and $d{\cal H}/dt=0$.
The random force {\bxi}$_{ij}$ 
acting on the particles $i$ and $j$ must be
orthogonal to their relative velocities 
${\bv}_{ij}$ and, therefore, performs no work.

To meet the condition (\ref{8}), we choose
two unit vectors ${\ba}_{ij}$ and ${\bb}_{ij}$,
to be orthogonal to ${\bv}_{ij}$, and to
each other,
\beq
{\bv}_{ij} \cdot {\ba}_{ij} = 0,
\qquad {\bv}_{ij} \cdot {\bb}_{ij} = 0,
\qquad {\bv}_{ij} \cdot {\bb}_{ij} = 0
\label{14}
\eeq
We also introduce two new scalar random variables
$\sigma_{ij}$ and $\eta_{ij}$, and write
\beq
{\bxi}_{ij} = {\ba}_{ij} \sigma_{ij} 
+ {\bb}_{ij} \eta_{ij}.
\label{60}
\eeq
Due to the orthogonal property (\ref{14}) of
${\ba}_{ij}$ and ${\bb}_{ij}$, we see that the
condition (\ref{8}) is fulfilled.

The vectors ${\ba}_{ij}$ and ${\bb}_{ij}$
are chose to be given by the expressions
(\ref{25a}) and (\ref{25b}). We remark that
when we interchange $i$ and $j$, the vector
${\ba}_{ij}$ preserves its sign and ${\bb}_{ij}$
changes sign. As ${\bxi}_{ij}$ must change sign
then the stochastic variable $\sigma_{ij}$ should change
its sign and $\eta_{ij}$ should preserve its sign.

The random variables 
$\sigma_{ij}$ and $\eta_{ij}$ are chosen to have
zero mean and the same variance $2\gamma_{ij}$,
which represents the strength of the
random forces and might depend on the positions of the
particles $i$ and $j$. It is reasonable to assume that
$\gamma_{ij}$ is nonzero only when the particles $i$
and $j$ are close to each other. 

The variance of ${\bxi}_{ij}$ is 
$\la{\bxi}_{ij}{\bxi}_{ij}\ra=2{\mathbb K}_{ij}$,
where ${\mathbb K}_{ij}$ is a $3\times3$ symmetric
matrix given by
\beq
{\mathbb K}_{ij} = \gamma_{ij}({\ba}_{ij}{\ba}_{ij}
+{\bb}_{ij}{\bb}_{ij}),
\label{18}
\eeq
and ${\mathbb K}_{ii}=0$.
The covariances of the random vector $\bzeta_i$ are
obtained from (\ref{7}) and is related to this matrix by
\beq
\la{\bzeta}_i {\bzeta}_j\ra = - 2{\mathbb K}_{ij},
\qquad i\neq j,
\label{22a}
\eeq
\beq
\la{\bzeta}_i {\bzeta}_i\ra = 2\sum_j {\mathbb K}_{ij}.
\label{22b}
\eeq

\subsection{Evolution equation}

To derive the equation that gives the time
evolution of the probability density function
$\rho$ of the dynamic variables ${\bx}_i$ and
${\bp}_i$, we assume the following discretized
equations of motion
\beq
\Delta{\bx}_i = \tau{\bv}_i,
\label{16a}
\eeq
\beq
\Delta{\bp}_i = \tau{\bf F}_i + \sqrt{\tau}{\bzeta}_i
+ \frac\tau4 \frac{\partial}{\partial{\bp}_i}
\cdot{\bzeta}_i {\bzeta}_i,
\label{16b}
\eeq
where ${\bzeta}_i$ is the sum of the independent 
stochastic variables ${\bxi}_{ij}$, each one 
given by (\ref{60}). 

We derive the evolution equation by the method that
we have used above. To this end we define the
characteristic function $\Phi$ by
\beq
\Phi = \int e^{i\Omega}\rho\, dqdp,
\eeq
where the integration is over the phase space and
\beq
\Omega = \sum_j ({\bk}_j\cdot {\bx}_j
+ {\bq}_j \cdot {\bp}_j),
\eeq
where ${\bk}_i$ and ${\bq}_i$ are vectors in the
Fourier space. The characteristic function is also the
average over $\rho$,
\beq
\Phi = \la e^{i\Omega} \ra.
\eeq

The finite variation $\Phi$ is given by
\beq
\Delta\Phi = \la e^{i\Omega}(e^{i\Delta{\Omega}}-1)\ra.
\eeq
Expanding the right-hand side of this equation 
up to terms of the order $\tau$, we find the
time derivative of $\Phi$ as
\[
\frac{\partial\Phi}{\partial t} = \sum_j\la e^{i\Omega}
(i{\bk}_j\cdot {\bv}_j + i{\bq}_j \cdot {\bf F}_j)\ra +
\]
\beq
+ \frac12 \sum_j 
\la e^{i\Omega}(\frac{\partial}{\partial{\bp}_j}
\cdot{\bzeta}_j {\bzeta}_j + \frac12\sum_{\ell}
i{\bq}_\ell \cdot {\bzeta}_\ell {\bzeta}_j)\cdot i{\bq}_j\ra.
\label{36}
\eeq

To obtain the evolution equation, it suffices to
take the inverse Fourier transform of the equation
(\ref{36}). To this end we observe that the first 
term of (\ref{36}) is the Fourier transform of
the Poisson brackets
\beq
\{{\cal H},\rho\} = 
\sum_j \left(\frac{\partial{\cal H}}{\partial {\bf x}_j}
\cdot\frac{\partial\rho}{\partial {\bf p}_j}
- \frac{\partial{\cal H}}{\partial {\bf p}_j}
\frac{\partial\rho}{\partial {\bf x}_j}\right),
\label{28}
\eeq
a result which is reached by using the relations
\beq
i{\bk}_j e^{i\Omega} = \frac{\partial e^{i\Omega}}
{\partial{\bx}_j},
\qquad 
i{\bq}_j e^{i\Omega} = \frac{\partial e^{i\Omega}}
{\partial{\bp}_j}.
\label{37}
\eeq

Using (\ref{22a}) and (\ref{22b}), the second term of
(\ref{36}) can be written in the form
\beq
\frac12 \sum_{j\neq\ell}\la e^{i\Omega}
(\frac{\partial}{\partial{\bp}_j}\cdot {\mathbb K}_{j\ell}
+ \frac{i}2({\bq}_j-{\bq}_\ell)\cdot {\mathbb K}_{j\ell})
\cdot i{\bq}_j\ra.
\eeq
Taking into account that ${\mathbb K}_{j\ell}$ depends
on ${\bp}_j$ through the difference
${\bp}_j-{\bp}_\ell$, we may write
\beq
\frac12 \sum_{j\neq\ell}\la e^{i\Omega}
({\bf D}_{j\ell} \cdot {\mathbb K}_{j\ell}
+ i({\bq}_j-{\bq}_\ell)\cdot {\mathbb K}_{j\ell})
\cdot i{\bq}_j\ra,
\eeq
where
\beq
{\bf D}_{j\ell} = \frac{\partial}{\partial{\bp}_j}
- \frac{\partial}{\partial{\bp}_\ell}.
\eeq
This expression is equal to the integral
\beq
- \frac12 \sum_{j\neq\ell}\int 
e^{i\Omega} i{\bq}_j \cdot {\mathbb K}_{j\ell}  
\cdot {\bf D}_{j\ell}\rho\,dxdp,
\eeq
which was obtained by an integration by parts
and using the result
\beq
{\bf D}_{j\ell} e^{i\Omega}
= i({\bq}_j-{\bq}_\ell)e^{i\Omega}.
\eeq
Employing again the second equality of (\ref{37})
and performing another integration by parts, we find
a result which is the Fourier transform of
\beq
\frac12 \sum_{j\neq\ell}
\frac{\partial}{\partial{\bp}_j} \cdot 
{\mathbb K}_{j\ell} \cdot {\bf D}_{j\ell}\rho.
\label{38}
\eeq

The evolution equation is obtained by observing
that $\partial\Phi/\partial t$ is the 
Fourier transform of $\partial\rho/\partial t$.
Therefore, to reach the equation it suffices to
add up the results (\ref{28}) and (\ref{38}),
\beq
\frac{\partial\rho}{\partial t} = 
\{{\cal H},\rho\} + 
\frac12 \sum_{(ij)}
{\bf D}_{ij} \cdot
{\mathbb K}_{ij} \cdot {\bf D}_{ij}\rho,
\label{26}
\eeq
Replacing ${\mathbb K}_{ij}$ in this expression,
the evolution equation acquires the form
\beq
\frac{\partial\rho}{\partial t} = 
\{{\cal H},\rho\} + \frac12 \sum_{(ij)}
\gamma_{ij} {\bf D}_{ij} \cdot 
({\ba}_{ij}A_{ij} +{\bb}_{ij} B_{ij}),
\label{26a}
\eeq
where
\beq
A_{ij} = {\ba}_{ij}\cdot {\bf D}_{ij}\rho,
\qquad
B_{ij} = {\bb}_{ij}\cdot {\bf D}_{ij}\rho.
\label{33}
\eeq

We have argued above in the analysis of the
stochastic motion of two particles that 
the kinetic energy is a stationary solution of the
evolution equation.  
The reasoning based on the Lagrange multipliers
can be generalized leading to the rule that a
{\it conserved quantity is
a stationary solution of the evolution equation}.
This is indeed the case of ${\cal H}$ and ${\bf P}$
in relation to the evolution equation (\ref{26a}),
which can be verified by inspection. The Poisson
brackets vanish trivially when $\rho$
is replaced by ${\cal H}$. When it is replaced by
${\bf P}$ it vanish as well if we use the condition
(\ref{19}). The quantities $A_{ij}$ and $B_{ij}$
given by (\ref{33}) vanish when $\rho$ is replaced
by ${\cal H}$ because ${\bf D}_{ij}{\cal H}$ equals
${\bv}_{ij}$ which is orthogonal to ${\ba}_{ij}$
and ${\bb}_{ij}$. The quantities $A_{ij}$ and
$B_{ij}$ also vanish when $\rho$ is replaced by
${\bf P}$ because ${\bf D}_{ij}\cdot {\bf P}=0$.

From the results just obtained, it follows
that any function of the conserved quantities
will be a stationary solution of the evolution
equation. In particular, the 
microcanonical Gibbs distribution (\ref{9})
is a stationary solution of the evolution equation,
and represents the 
state of thermodynamic equilibrium of the system.
If we start with a probability distribution $\rho$
defined on a surface of constant energy, then it will
remain on this surface. If the energy is
equal to $E$, the probability density will
eventually reach the Gibbs distribution (\ref{9}).
In other words the system will relax to equilibrium.
This result can be demonstrated by showing that
the entropy is a nondecreasing function of time
and that its maximum value is the one corresponding
to the Gibbs distribution (\ref{9}). This will 
shown in the following.

\subsection{Entropy production}

Let us determine the time evolution of the average
$U$ of ${\cal H}$. Multiplying equation (\ref{26}) by
${\cal H}$ and integrating in the phase space,
we arrive at the equation
\beq
\frac{dU}{dt} = - \frac12 \sum_{(ij)}\int
{\bv}_{ij} \cdot {\mathbb K}_{ij}
\cdot {\bf D}_{ij}\rho\, dxdp,
\label{34}
\eeq
where we have performed an integration by parts
and used the result ${\bf D}_{ij}{\cal H}={\bv}_{ij}$.
Taking into account the orthogonality (\ref{14}),
it follows that the right-hand side of (\ref{34})
vanishes and $dU/dt=0$. That is, $U$ remains constant
in time, which is expected because ${\cal H}$ is 
conserved by the evolution equation.

We determine now the time evolution of 
the entropy $S$, defined by
\beq
S = -k \int \rho \ln\rho \, dxdp.
\label{48}
\eeq
Multiplying the evolution equation (\ref{26}) by
$-\ln\rho$ and integrating in the phase space, we find
\beq
\frac{dS}{dt} = \frac{k}2 \sum_{(ij)} \int
\frac1\rho({\bf D}_{ij}\rho) \cdot
{\mathbb K}_{ij} \cdot ({\bf D}_{ij}\rho),
\eeq
where we have performed an integration by parts.
Replacing ${\mathbb K}_{ij}$ given by (\ref{18}),
we reach the following expression
\beq
\frac{dS}{dt} = \frac{k}2 \sum_{(ij)} \int 
\frac{\gamma_{ij}}\rho (A_{ij}^2 + B_{ij}^2)dxdp.
\label{29}
\eeq

We see that the right-hand side of (\ref{29}),
which we denote by $\Pi$, is positive definite
and should be interpreted as the rate of entropy
production since there is no exchange 
of energy with the environment, in fact no exchange
of heat since there is no external work involved.
We remark that $\Pi$ vanishes in equilibrium
because each one of the terms $A_{ij}$ and $B_{ij}$
vanish when $\rho$ is replaced by the equilibrium 
distribution $\rho_e$, given by (\ref{9}).
Since $S$ is monotonically increasing in time because
$dS/dt=\Pi\geq0$, it reaches a maximum when 
$\Pi=0$, that is, when the distribution is $\rho_e$.

\section{Liouville equation}

If we set $\gamma_{ij}$ equal to zero in the
Fokker-Planck equation (\ref{26}), which means
that no stochastic forces are involved,
it reduces to the form
\beq
\frac{\partial\rho}{\partial t} + \{\rho,{\cal H}\} = 0,
\label{3}
\eeq
which is the Liouville equation. Here we present
the usual derivation of the Liouville equation,
which is carried out by the use of the Liouville
theorem. We also present some of its properties
in order to compare with those of the
Fokker-Planck equation (\ref{26}). We consider
a system described by a Hamiltonian ${\cal H}$,
and the equations of motion are
\beq
\frac{dx_i}{dt} = \frac{\partial{\cal H}}{\partial p_i},
\qquad 
\frac{dp_i}{dt} = - \frac{\partial{\cal H}}{\partial x_i}.
\eeq

If at time $t$ we consider a region ${\cal R}$
in phase space which is transformed, through
the Hamiltonian motion, in a region ${\cal R}'$
at time $t'$, then the Liouville theorem states
that the volume of these regions are equal, that is,
\beq
\int_{\cal R} dx dp = \int_{\cal R'} dx' dp'.
\eeq
We may also write this result as $dxdp = dx'dp'$
which means that the Jacobian of the transformation
$(x,p)\to(x',p')$ equals the unity.

In the usual derivation, the Liouville equation is
obtained by postulating that
the probability of the region ${\cal R}$ at time
$t$ is equal to that of the region ${\cal R'}$ at
time $t'$, that is, 
\beq
\int_{\cal R} \rho(x,p,t) dx dp
= \int_{\cal R'} \rho(x',p',t')dx' dp',
\eeq
or $\rho(x,p,t) dx dp = \rho(x',p',t')dx' dp'$.
From this equation and the Liouville theorem,
we find
\beq
\rho(x,p,t)=\rho(x',p',t'),
\label{4}
\eeq
and the probability density is invariant along a
trajectory. This is equivalent to say that the
material time derivative of $\rho$ vanishes, that is,
\beq
\frac{\partial\rho}{\partial t} +
\sum_i \left(\frac{\partial\rho}{\partial x_i}
\frac{\partial x_i}{dt}
+ \frac{\partial\rho}{\partial p_i}
\frac{\partial p_i}{dt}\right) = 0.
\eeq
Taking into account the Hamilton equations of
motion, the second term becomes the Poisson brackets
$\{\rho,{\cal H}\}$ and we arrive at the Liouville
equation (\ref{3}).

The most relevant property of the Liouville equation
is that the microcanonical Gibbs distribution (\ref{9})
is a stationary distribution of the equation. However,
not all initial distribution will relax to this
distribution. To discuss this point we use the
property coming from the Liouville equation
that the entropy is invariant in time. 

The entropy $S(t)$ at time $t$ is defined by
(\ref{48}). Using the invariance of
the probability density along a trajectory,
given by equation (\ref{4}), we reach the
result $S(t)=S(t')$, at distinct instants of time.
Alternatively, we may employ the procedure used
above in the stochastic approach.
We multiply the Liouville equation (\ref{3})
by $k\ln\rho$ and integrate in the phase
space. After an integration by parts we find
that $dS/dt=0$, and the entropy is invariant.

To proceed in our analysis it is convenient
to define the measure of a surface A of 
constant energy $E$ in phase space. We consider
another surface A$'$ of constant energy $E'$,
with $E'$ differing little from $E$.
The measure of A, which we denote by $\Omega_{\rm A}$,
is the volume between the two surfaces A$'$ and A
divided by $E'-E$. Thus $\Omega$ appearing in equation
(\ref{9}) is the measure of the whole surface of
constant energy $E$. 

Let us consider a probability distribution,
denoted by $\rho_{\rm A}$, which is nonzero only
in a surface $A$ which is a subset of the
whole surface of constant energy $E$. Within
$A$ it is given by
\beq
\rho_{\rm A} = \frac1\Omega_{\rm A}\delta(E-{\cal H}).
\label{9a}
\eeq
where $\Omega_{\rm A}$ is the measure of A.
The corresponding entropy is $S_{\rm A}=k\ln\Omega_{\rm A}$.

Suppose that the initial condition to the Liouville
equation is $\rho_{\rm A}$. As time evolves, the
surface $A$ changes to another surface A$'$ and
become intertwined with the complementary surface
to the whole surface. However, by the Liouville
theorem, the measure of A$'$ is the same as that
of A, and the entropy remains constant and equal
to $S_{\rm A}$. Therefore the Gibbs distribution
will never be reached because the entropy of this
distribution is $S_e=k\ln\Omega$ which is larger
than $S_{\rm A}$.

We present now an example where the initial
distribution is nonzero and given by
\beq
\rho = a \rho_{\rm A} + b \rho_{\rm B},
\label{10}
\eeq
where $a+b=1$, and the surfaces A and B are
complementary surfaces and make up the whole
surface of constant energy $E$. 
The corresponding entropy is
\beq
S = k a \ln\frac{\Omega_{\rm A}}{a} +
k b \ln\frac{\Omega_{\rm B}}{b},
\eeq
and it differs from $S_e=k\ln\Omega$,
where $\Omega=\Omega_{\rm A}+ \Omega_{\rm B}$. 
In fact we can show that $S_e\geq S$,
the equality occurring when $\rho$ is the
Gibbs distribution in which case
$a=\Omega_{\rm A}/\Omega$ and
$b=\Omega_{\rm B}/\Omega$.
Therefore, if the initial distribution is
of the type (\ref{10}) but is not the
Gibbs distribution, this Gibbs distribution 
will never be reached.

\section{Conclusion}

We have proposed a stochastic equation of motion
that emerges as a consequence of stochastic forces
acting on each pair of particles. The random forces
acting on a pair of particles are action and reaction
leading to the conservation of momentum. These
forces are perpendicular to the relative
velocity between the particles, performing 
therefore zero work on the two particles,
and thus preserving their kinetic energy.

These stochastic forces and the conservative forces
leads to the conservation of energy along
a trajectory in phase space. The trajectories
in phase space are restricted to the surface
of constant energy and is thus similar in this
aspect to the Hamiltonian flow. From the stochastic
motion, we have derived the evolution equation
for the density which turns out to be a 
Fokker-Planck equation, but distinct from 
the usual Fokker-Planck equation that describes
the contact of system with a heat reservoir.

To derive the evolution equation 
we introduced an appropriate interpretation of
the stochastic equation of motion that does
not correspond to the interpretation
proposed by Itô nor that proposed by Stratonovich.
The interpretation proposed here makes the 
energy to be conserved at all times, a result that
was demonstrated by showing that the
Hamiltonian function is a stationary solution 
of the evolution equation.

In contrast to the Liouville equation, which
also describes a closed system and conserves
the energy, the
evolution equation that we have set up predicts
the increase of entropy and the relaxation to
equilibrium. In other terms, the equation
describes an irreversible decay to equilibrium
but we cannot acertain whether the present approach
could describe correctly the actual decay observed in
real closed systems.


\end{document}